\documentclass[12pt]{article} 
\usepackage{graphicx}

\usepackage{pazha}
\tightenlines

\def\*{$^{*}$}
\def\a{$^{\mbox{\small a}}$}
\def\b{$^{\mbox{\small b}}$}
\def\c{$^{\mbox{\small c}}$}
\def\d{$^{\mbox{\small d}}$}
\def\e{$^{\mbox{\small e}}$}

\def\INTEGRAL{\hbox{INTEGRAL}}
\def\sax{\hbox{SAX\,J1818.6-1703}}
\def\ax{\hbox{AX\,J1749.1-2733}}
\def\deg{$^\circ$}
\def\etal{{\it et~al.}}

\sloppypar

\begin{document}
\begin{flushright}
{\it to be published in Astronomy Letters, v. 31, n. 10,
  pp. 672--680 (2005)}
\end{flushright}

\vspace{3cm} 

\baselineskip 21pt

\title{\bf OUTBURST OF THE X-RAY TRANSIENT \sax\ DETECTED BY 
  {\sl INTEGRAL}\/ IN SEPTEMBER 2003}

\author{\bf 
S. A. Grebenev\affilmark{1*} and R. A. Sunyaev\affilmark{1,2}}

\affil{
{\it Space Research Institute, Russian Academy of Sciences,
  Russia}$^1$\\ 
{\it Max-Planck-Institut f\"{u}r Astrophysik,
  Garching, Germany}$^2$}

\vspace{2mm}
\received{May 16, 2005}

\sloppypar 
\vspace{2mm}
\noindent
During the observation of the Galactic-center field by the
\INTEGRAL\ observatory on September 9, 2003, the
\mbox{IBIS/ISGRI} gamma-ray telescope detected a short
(several-hours-long) intense ($\sim 380$ mCrab at the peak)
outburst of hard radiation from the X-ray transient
\sax. Previously, this source was observed only once in 1998
during a similar short outburst. We present the results of our
localization, spectral and timing analyses of the object and
briefly discuss the possible causes of the outburst. The release
time of the bulk of the energy in such an outburst is appreciably
shorter than the accretion (viscous) time that characterizes the
flow of matter through a standard accretion disk.


\vfill
\noindent\rule{8cm}{1pt}\\
{$^*$ E-mail $<$sergei@hea.iki.rssi.ru$>$}

\clearpage

\section*{INTRODUCTION}
\noindent
The source \sax\ was discovered by the BeppoSAX observatory on
March 11, 1998, during an X-ray outburst that lasted only a few
hours (in't Zand \etal\ 1998; in't Zand 2001). The appearance of
a new transient near the well-known burster GX\,13+1 (at an
angular distance of $\sim1$\deg) was recorded by the WFC-2
wide-field X-ray camera at 19\uh10\um\ (UT). By 20\uh40\um, the
photon flux from \sax\ reached its maximum: $\sim100$ mCrab in
the range 2--9 keV and $\sim400$ mCrab in the range 9--25
keV. The observation was interrupted 3 h later, but the flux was
almost halved by this time, indicative of a fast decay of the
transient. The source's position, $R.A. = $ 18\uh18\um39\us,
$Decl. = -$17\deg 03\farcm1 (epoch 2000.0), was determined with
an accuracy of 3\arcmin. Only one catalogued B3\,III star,
HD\,168078, with $V=10\fm7$ is within the error circle (in't
Zand \etal\ 1998), but there is no additional reasons to believe
it to be a real candidate for optical identification with \sax.

All that has been known about \sax\ until recently is listed
above. Its nature and, primarily, the mechanism that produced
such a short outburst with a duration much shorter than the
characteristic time scale for the propagation of disturbances in
a standard accretion disk ($t_{\rm vis} \ga 1.4$ days) have
remained unclear. In this paper, we present the results of our
observations of the second outburst of hard radiation from this
source that has allowed us to investigate it in more
detail. This flare was detected on September 9, 2003, by the
\INTEGRAL\ observatory.

\section*{OBSERVATIONS}
\noindent
The \INTEGRAL\ international gamma-ray astrophysics observatory
(Winkler \etal\ 2003) was placed in a high apogee orbit by a
PROTON launcher on October 17, 2002 (Eismont \etal\ 2003). There
are four telescopes on its board that allow concurrent X-ray,
gamma-ray, and optical observations of cosmic sources. This work
is based on the data obtained with the IBIS gamma-ray telescope
at energies above 18 keV. Unfortunately, no concurrent
observations of \sax\ were performed in the standard X-ray range
(by the \mbox{JEM-X} telescope). Since the source is fairly far
from the Galactic center toward which the observatory was
pointed, it was not within the field of view of the \mbox{JEM-X}
telescope, which is narrower than that of the \mbox{IBIS}
telescope.

The \mbox{IBIS} telescope (Ubertini \etal\ 2003) uses the
principle of a coded aperture to image the sky in hard X-rays and
gamma-rays in a 30\deg $\times$30\deg\ field of view (the fully
coded field is 9\deg $\times$9\deg) with an angular
resolution of 12\arcmin\ (FWHM). It is equipped with 
two position-sensitive detectors: ISGRI (Lebrun \etal\ 2003)
composed of 128$\times$128 CdTe semiconductor elements with a high 
sensitivity in the range 18-200 keV and PICsIT located under it
(Labanti \etal\ 2003) and composed of 64$\times$64
scintillators CsI(Tl) with an optimal sensitivity in the range 
175 keV -- 10 MeV. In this paper, we use only the ISGRI data.
The total area of the sensitive elements of this detector is
2620 cm$^2$; the effective area for sources at the center of the
field of view is $\sim 1100$ cm$^2$ (half of the detector is
shadowed by opaque aperture elements). The detector provides
fairly good energy, $\Delta E/E\sim 7$\% (FWHM), and high time,
$\Delta t\simeq 61\ \mu$s, resolutions.

Although the outburst of \sax\ was initially detected using a
standard software package for analyzing the \INTEGRAL\ data (at
that time, OSA-3.1), all of our results presented in this paper
were obtained using the data processing codes developed for the
IBIS/ISGRI telescope at the Space Research Institute of the
Russian Academy of Sciences. A general description of the
procedures used can be found in a paper by Revnivtsev \etal\
(2004). Application of the latest version of these programs to
the observations of the Crab Nebula has shown that the
systematic measurement error of the absolute photon flux from
the source does not exceed 7\%, while the measurement error of
the relative fluxes in various spectral channels does not exceed
3\%. In our spectral analysis, we used the response matrix of
the OSA-4.2 standard package (rmf-file of version 12 and the
arf-file of version 6), which proved to be good at fitting the
spectra of the Crab Nebula, in particular, the spectra measured
in August 2003 immediately before the observations under
consideration. The spectrum of the Nebula was assumed to be
$dN(E)/dE=10\, E^{-2.1}$ phot cm$^{-2}$ s$^{-1}$ keV$^{-1}$,
where the energy $E$ is given in keV. For our study within the
framework of the same general approach to analyzing the
IBIS/ISGRI data, we developed codes for reconstructing the
source's light curves with a good time resolution.

When the outburst of the transient \sax\ occurred, the \INTEGRAL\
observatory was performing a long (with a total exposure time
of 2 Ms) series of observations of the Galactic-center
region. During this series, the IBIS telescope detected 60 hard
radiation sources of various intensities (Revnivtsev \etal\
2004). Curiously enough, another poorly studied source,
\ax, flared up in this region on September 9, 2003,
almost simultaneously with \sax. The results of its study are
presented elsewhere (Grebenev and Sunyaev 2005). The  
INTEGRAL observations in this period were performed by
successive pointings at points of the Galactic-center field
spaced $\sim 2$\deg\ apart according to the $5\times5$
scheme. Depending on the pointing, the exposure efficiency of
\sax\ changed greatly. The duration of each pointing was
$\sim3450$ s. 

\section*{RESULTS}
\noindent
The appearance of the transient source in the field of view was
first recorded at a statistically significant level by the
IBIS/ISGRI telescope (at a signal-to-noise ratio of $S/N=9.4$)
during the pointing that began on September 9, 2003, at
00\uh01\um\ (UT). Analysis of its X-ray image showed that \sax\
flared up. The measured photon flux from it was $69\pm7$ mCrab
in the range 18--45 keV and $43\pm16$ mCrab in the range 45--70
keV (the flux of 1 mCrab in these ranges corresponds to
radiation fluxes of $1.1\times10^{-11}$ and $4.7\times10^{-12}$
erg cm$^{-2}$ s$^{-1}$, respectively). The brightness of the
source remained the same during the next pointing; subsequently,
it faded for 2--3 h\footnote{Since the pointings at this time
were particularly unfavorable for the source's observations (it
was at the very edge of the field of view), the flux was
measured with large errors; nevertheless, the fall in flux
appears statistically significant (see Fig.~1).}, but flared up
again. During the pointings that began at 10\uh42\um\ and
12\uh41\um, two superintense bursts were detected from the
source during which it became the brightest among all of the
sources in the field of view. During the first (stronger) burst,
the measured photon flux in the above ranges reached $230\pm5$
and $172\pm10$ mCrab, respectively. The source remained
moderately bright ($\sim 50$--$70$ mCrab) until 20\uh, we failed
to detect it in several subsequent pointings, the observations
were then interrupted, because the satellite entered the Earth's
radiation belts in the final segment of its orbit. During the
next orbital cycle (September 10--13), the source was detected
only at the telescope's sensitivity limit ($S/N\simeq6.0$) with
a mean 18--45-keV flux of $6.6\pm1.1$ mCrab.

\begin{figure}[ht]
\hspace{-0.5cm}\includegraphics[width=17cm]{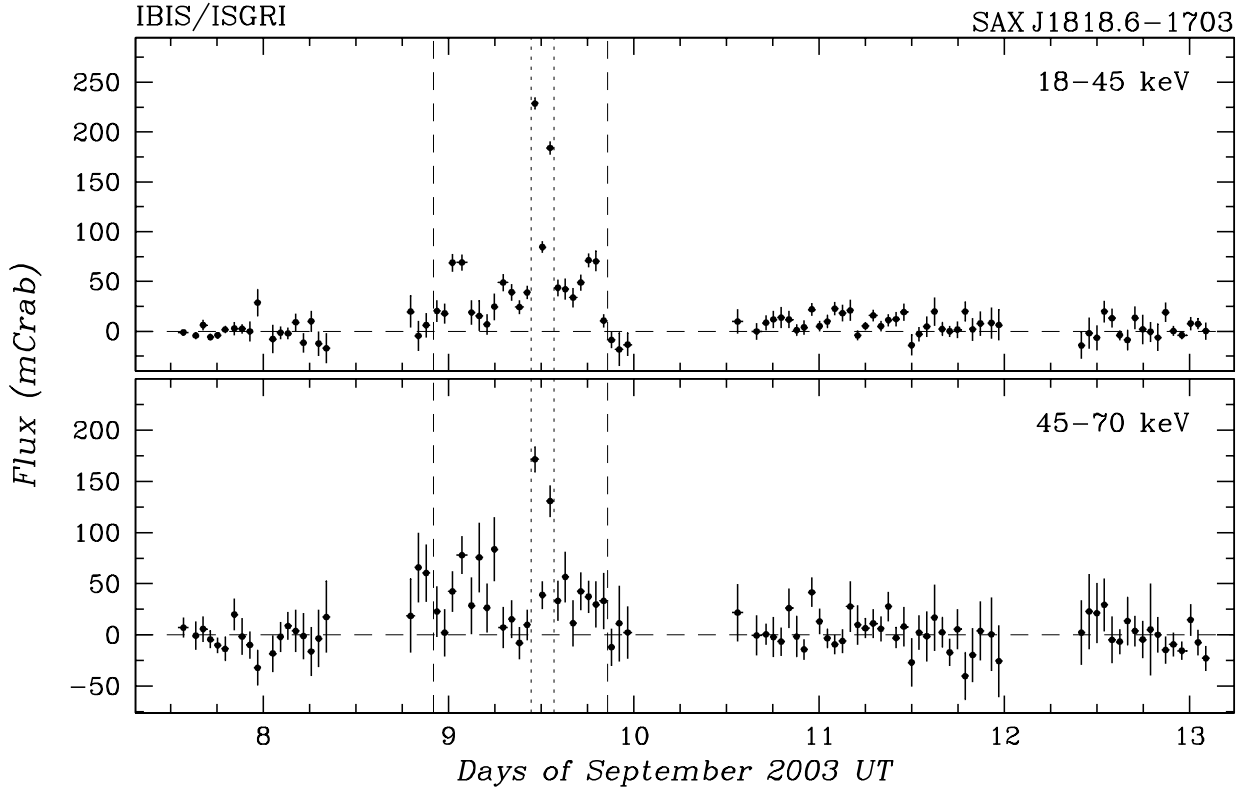}

\caption{\rm IBIS/ISGRI light curve for \sax\ in the energy
  ranges 18--45 and 45--70 keV obtained in the period September
  7-13, 2003. Each point of this curve corresponds to an
  individual $\sim3450$-s-long pointing of the \INTEGRAL\
  observatory. The dashed and dotted vertical lines indicate the
  source's activity period and the main event of the outburst
  (two intense bursts of hard radiation).}
\end{figure}

The described picture is illustrated by Fig.~1, which shows the
source's light curves constructed from the observations of  
September 7--13 in two energy ranges. Each point of these curves
is the measurement of the photon flux from the source in the  
corresponding sky image obtained during an individual
pointing. 

The vertical dashed lines in Fig.~1 indicate the interval of the
source's statistically significant detection. We used this
interval to accumulate the integrated images of the sky near
\sax\ (signal-to-noise maps) in the energy ranges 18--45 and
45--70 keV shown in Figs.~2 and 3, respectively. The total
exposure time was 75400~s. Apart from \sax, four more X-ray 
sources are seen in Fig.~2: the bursters GX\,17+2 and GX\,13+1,
the ``atoll'' source GX\,9+1, and the X-ray pulsar
\mbox{IGR\,J18027-2016} (also known as
\mbox{SAX\,J1802.7-2017}). \sax\ is the brightest of these
sources --- it was detected at a signal-to-noise ratio of
$S/N\simeq48$, while the next brightest source GX\,9+1 has
$S/N\simeq20$. In Fig.~3, \sax\ is the only source (it is seen
at $S/N\simeq15$). Estimates indicate that this is attributable
not just to the natural decrease in the hard X-ray photon flux
typical of all sources; the flux from \sax\ falls in this case
more slowly, i.e., the source has a much harder (with the
possible exception of the X-ray pulsar \mbox{IGR\,J18027-2016})
spectrum. 

\begin{figure}[h]
\hspace{0.5cm}\includegraphics[width=15cm]{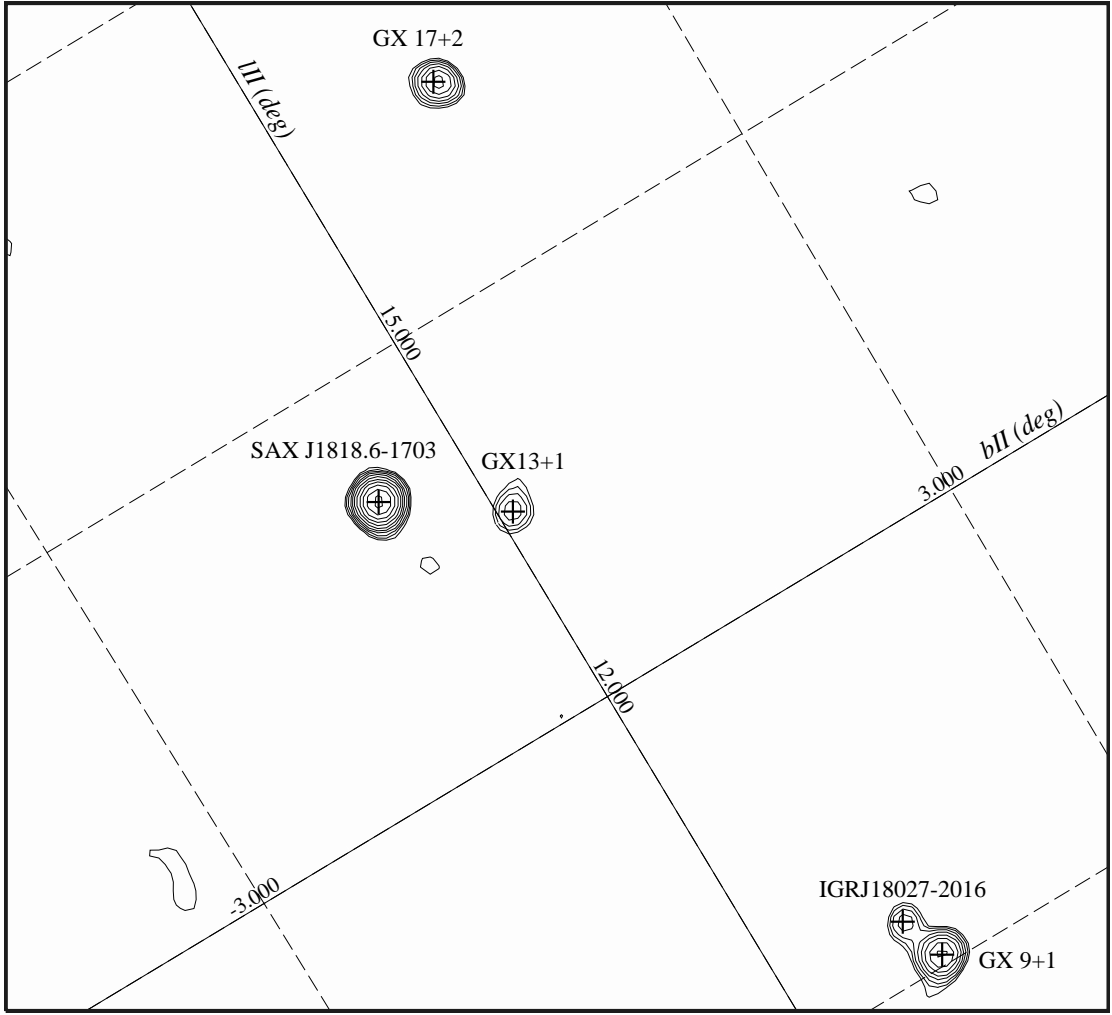}

\caption{\rm IBIS/ISGRI X-ray image of the region around \sax\
  obtained on September 9, 2003, during its outburst. The
  contours indicate the regions of confident detection of
  sources in the energy range 18--45 keV and are given at
  signal-to-noise ratios of 3, 3.9, 5.2, 6.8, 8.9, 11.6, ... (a
  logarithmic scale is used). The image size is approximately
  equal to 9\deg$\times9$\deg.}
\end{figure}
\begin{figure}[h]
\hspace{0.5cm}\includegraphics[width=15cm]{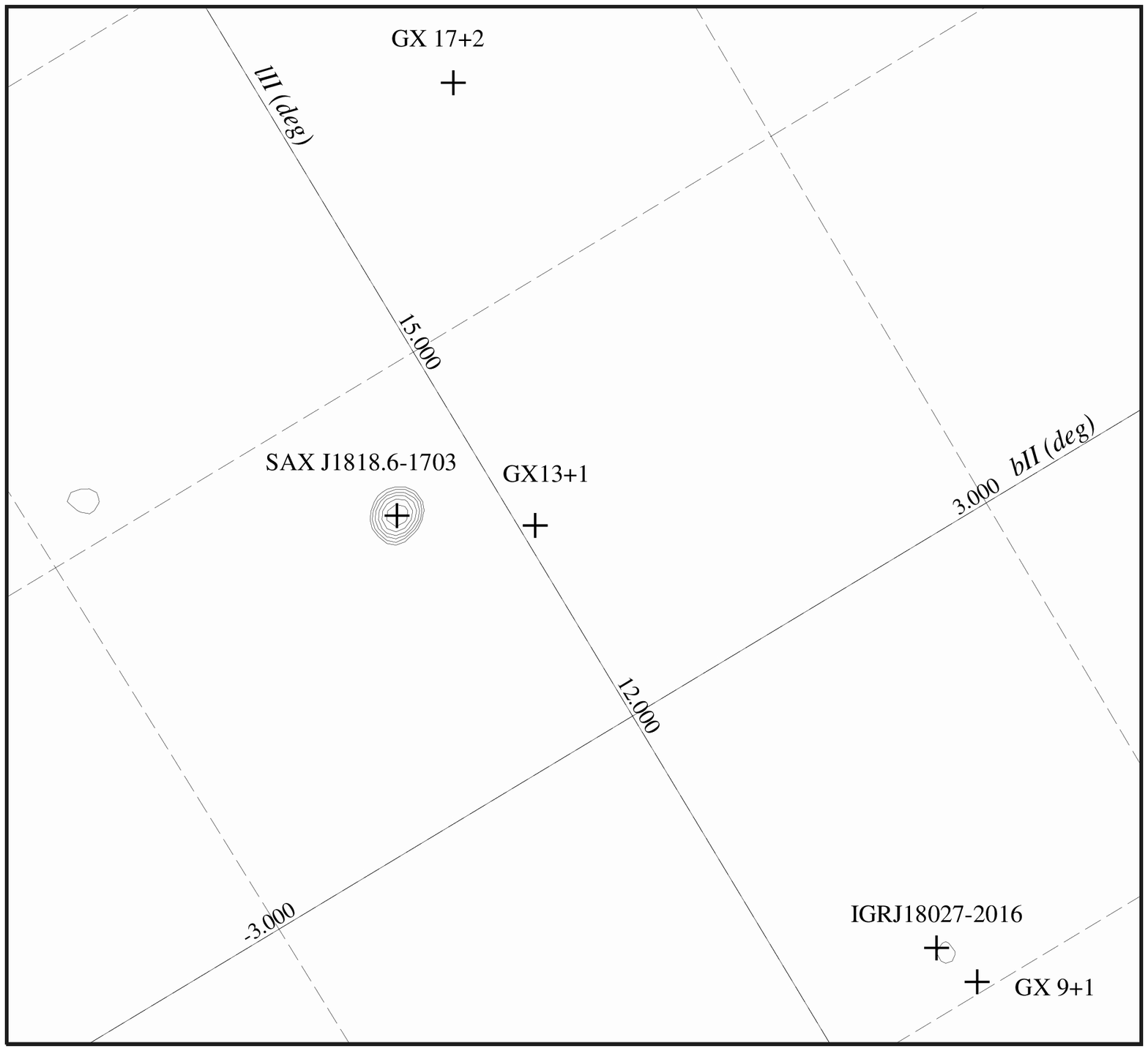}

\caption{\rm The same image as that in Fig.~2 but in the energy
  range 45--70 keV. Of the five sources observed in this field
  at low energies, only \sax\ is still bright (seen at $S/N
  \simeq 14.8\sigma$).}
\end{figure}
The image shown in Fig.~2 was used to improve the localization
of the source. The position found, $R.A. = 18$\uh18\um38\,\fs2,
$Decl. = -$17\deg 03\arcmin 11\arcsec\ (epoch 2000.0,
1.5\arcmin\ uncertainty), coincided with its position measured
by BeppoSAX in 1998 to within 12\arcsec.

\subsection*{The Outburst Time Profile}
\noindent
To elucidate the nature of the source's outburst, it is
crucially important to analyze the structure of the two intense
short bursts occurred at about 11 and 13 h. Since the light
curve in Fig.~1 does not allow this to be done, we reconstructed
more detailed light curves. The top panel in Fig.\,4 shows the
18--45-keV light curve with a time resolution (bin size) of
500~s. It spans only the source's activity period. The actually
measured count rate, i.e., the count rate corrected for the dead
time and other instrumental effects, but uncorrected for the
variations in the effective (source-irradiated) area of the
detector due to the change in the INTEGRAL pointing, is along
the Y axis. This effect is important, since the source was
outside the fully coded field of view of the telescope. We do
not make the corresponding correction in order not to overload
the figure. At the time resolution used, statistically
insignificant spikes appear in the corrected light curve due to
the Poissonian fluctuations of the count rate when the source
approaches the edge of the field of view. Instead, the bottom
panel in Fig.~4 shows the curve of variations in the effective
area of the detector. We clearly see its correlation with the
count rate, which, however, does not distort severely the main
event. The effective area for the observation of \sax\ was only
620~cm$^2$ even at maximum, i.e., it was almost a factor of $2$
smaller than the area typical of the sources in the fully coded
field of view. For such an effective area, 1 count/s
corresponded to a flux of $\sim 16.4$ mCrab in the energy range
under consideration, so, according to this figure, the maximum
flux from the source reached $\sim 380$ mCrab.

\begin{figure}[h]
\hspace{-0.5cm}\includegraphics[width=17cm]{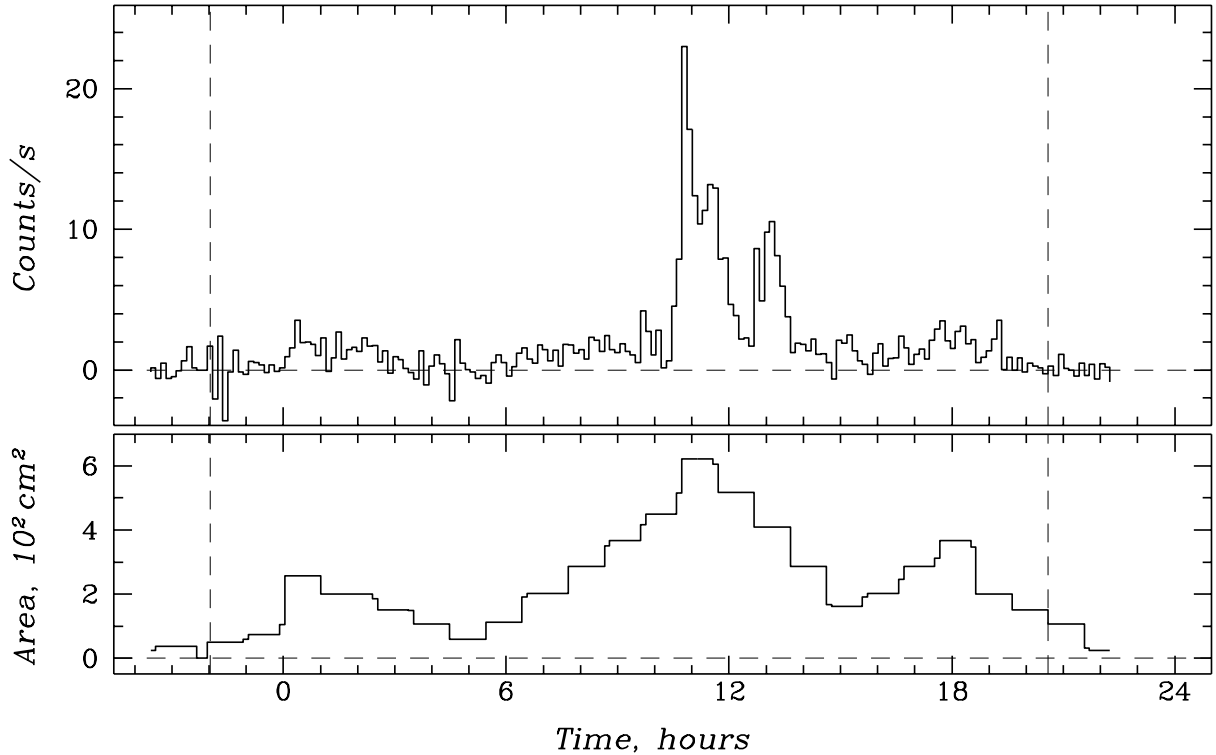}

\caption{\rm Detailed IBIS/ISGRI 18--45-keV light curve for
  \sax\ obtained on September 9, 2003, in the period of its
  activity (the top panel). The time in hours from the beginning
  of the day (UT) is along the X axis. The time resolution is
  500~s. The count rate along the Y axis was corrected for the
  dead time of the detector and other instrumental effects,
  except the variations in the effective area for this source
  related to the change in the \INTEGRAL\ pointing (the effect
  of partial coding). The corresponding change in the effective
  area is shown in the bottom panel.}
\end{figure}
\begin{figure}[p]
\hspace{3.5cm}\includegraphics[width=9cm]{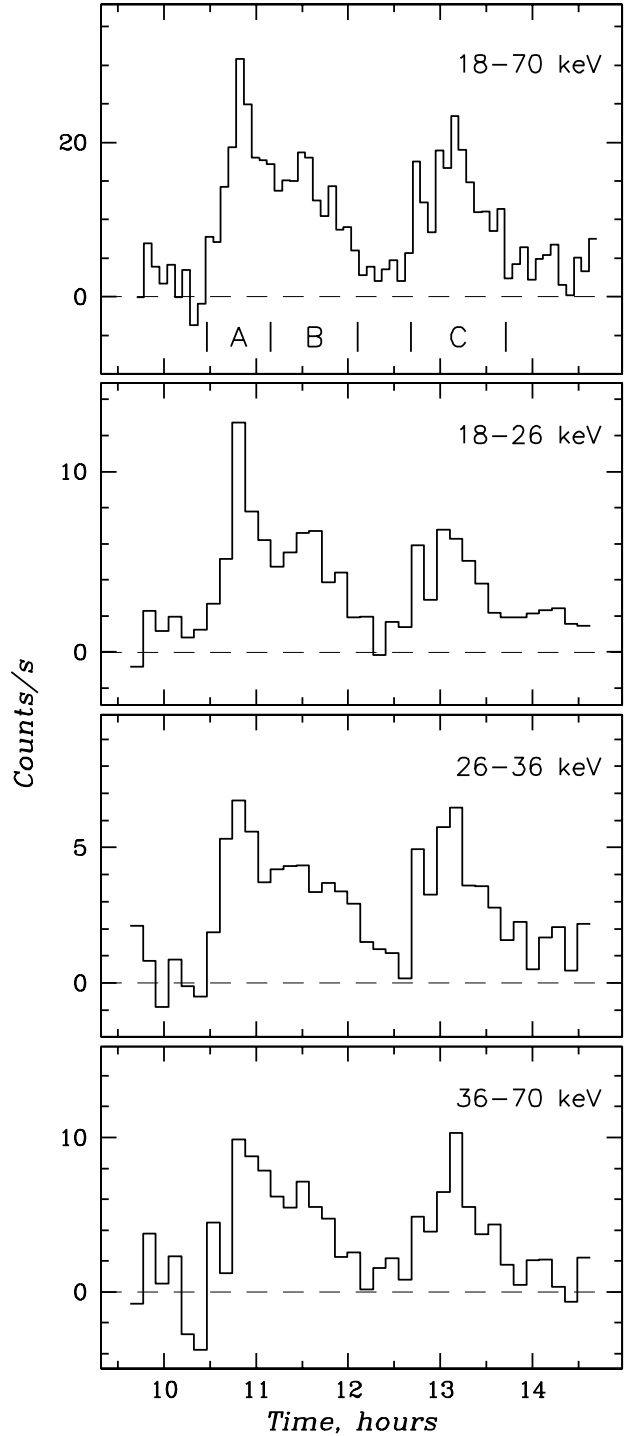}

\caption{\rm Change with energy of the time profile for the main
  outburst event of \sax\ recorded by the IBIS/ISGRI telescope
  on September 9, 2003. The resolution is 500~s everywhere,
  except the profile measured in a wide energy range, 18--70 keV
  (the bin size for it is 300~s). The count rate along the Y
  axis was corrected for all instrumental effects and reduced to
  the same area of 620 cm$^2$, which corresponds to the maximum
  achieved effciency of the source's observations (1 count/s is
  approximately equal to 36, 48, 31, and 12 mCrab in the ranges
  18--26, 26--36, 36--70, and 18--70 keV, respectively). The
  time in hours from the beginning of the day (UT) is along the
  X axis.}
\end{figure}

The figure suggests that the two bursts of the main outburst
event have a fairly complex time profile. A narrow (10--20 min)
precursor peak and a broader (1.5--2 h) main peak can be
distinguished in each of them. The amplitude of the precursor
peak in the first burst is almost twice that of the main peak,
while the precursor peak in the second burst is appreciably
smaller. In general, the two bursts resemble ordinary type-I
X-ray bursts with photospheric expansion, i.e., bursts produced
by a thermonuclear explosion on the neutron star surface at
which the photospheric luminosity reached the Eddington limit
(see Lewin \etal\ 1993). However, the duration of the bursts
from \sax\ was 2--3 h, which is much longer than the duration of
ordinary X-ray bursts. The recently discovered superbursts
(Kuulkers \etal\ 2002; in't Zand \etal\ 2004) have comparable
durations, but exhibit completely different time profiles --- a
very fast rise and a long exponential decay. As we will see
below, there are also more fundamental differences between these
bursts.

Figure~5 shows the time profiles of the main event in several
energy ranges after their reduction to the effective area of
620~cm$^2$. In general, they are similar in structure,
suggesting that the spectral shape of the source changed little
during the outburst. However, note a clear decrease with energy
in the relative amplitude of the precursor peak in the first
burst and a probable decrease in the amplitude of the precursor
peak in the second burst. These changes were confirmed during a
detailed spectral analysis.

\begin{figure}[ht]
\hspace{1.5cm}\includegraphics[width=12cm]{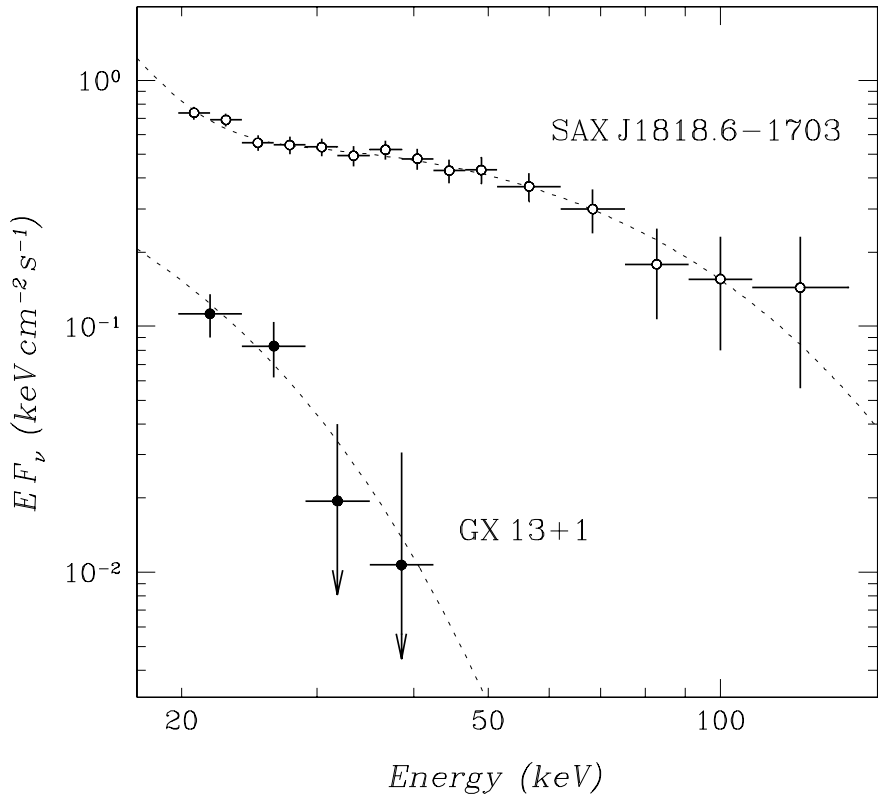}

\caption{\rm Average IBIS/ISGRI spectrum of \sax\ obtained on
  September 9, 2003, in the period of its activity (open
  circles). The spectrum is very hard; the characteristic
  temperature when fitting the spectrum by the bremsstrahlung
  law of an optically thin thermal plasma is $kT\simeq 36$
  keV. For comparison, the filled circles indicate the X-ray
  spectrum of the source GX13+1 closest to \sax\ measured at the
  same time, which is typical of accreting neutron stars with a
  weak magnetic field ($kT\simeq 5$ keV). At low energies, an
  additional soft radiation component is apparently present in
  the spectrum of \sax. The dotted lines indicate the best fits
  to the spectra (see the text).}
\end{figure}
\subsection*{The Radiation Spectrum}
\noindent
Figure~6 shows the average spectrum of \sax, obtained in the
period of its activity (during the interval bounded by the
vertical dashed lines in Figs.~1 and 4). The source's radiation
is recorded up to $\sim 200$ keV, with an exponential cutoff
being observed at energies above 70 keV. Note that an additional
soft ($h\nu<30$ keV) radiation component is present in the
spectrum. The energy characteristics of the source's radiation
in this period, its mean luminosity and energy release
calculated by assuming that \sax\ is actually near the Galactic
center, at a distance of $d\simeq8$ kpc\footnote{For the source
GX\,13+1 closest to \sax\ $d\simeq7\pm1$ kpc
(Bandyopadhyay \etal\ 1999).}, are given in Table~1.

\begin{table}[ht]
\vspace{6mm}
\centering
{{\bf Table 1}. Parameters of the outburst of
  \mbox{SAX\,J1818.6-1703}\protect\\
   observed with IBIS/ISGRI on September 9, 2003.}\label{meanpar}  

\vspace{5mm}\begin{tabular}{l|c|c} \hline\hline
Interval& Parameter\a\ &Value\\ \hline
Entire activity&$\Delta T$    & 22.5 h\\
period\b\ &$L_{\rm X}$  &$7.5\times10^{36}\ \mbox{erg c}^{-1}$\\
           &$F_{\rm X}   $  &$6.1\times10^{41}\ \mbox{erg}$\\ \hline
Main event\c\ &$\Delta T$      &2.7 h\\
     &$L_{\rm X}$     &$1.9\times10^{37}\ \mbox{erg s}^{-1}$ \\
           &$F_{\rm X}$     &$1.8\times10^{41}\ \mbox{erg}$\\ \hline
\multicolumn{3}{l}{}\\ [-3mm]
\multicolumn{3}{l}{\a\ Duration $\Delta T$, luminosity
  $L_{\rm X}$, and energy release  $F_{\rm X}$}\\
\multicolumn{3}{l}{\ \ \ in the range 20--200 keV for an assumed
  distance of 8 kpc}\\ 
\multicolumn{3}{l}{\b\ Bounded by the vertical dashed lines in Fig.\,1}\\
\multicolumn{3}{l}{\c\ The sum of intervals A, B, C marked in Fig.\,5}\\
\end{tabular}
\end{table}

Table 2 summarizes the results of fitting the spectrum by simple
analytical models: a power law (PL), a power law with an
exponetial high energy cutoff (PE), the radiation formed through
Comptonization of low energy photons in a cloud of high
temperature plasma (ST, Sunyaev and Titarchuk 1980), the
bremsstrahlung of an optically thin thermal plasma (TB), the
bremsstrahlung with an additional soft blackbody component
(TB+BB), and the Comptonization radiation with an additional
blackbody component (ST+BB). We see that even the
single-component PL, PE, and ST models describe satisfactorily
the radiation spectrum. Although the introduction of an
additional soft radiation component affects only the first two
or three points of the spectrum, this leads to further
significant improvement of its approximation. Extrapolating the
soft component to the X-ray energy range we get a luminosity
an order of magnitude higher than that in the hard energy
range. Thus, the overall energetics of the source could be
maintained at a level approaching the critical Eddington level
for accretion onto a neutron star (or, given the uncertainty in
the spectral shape of the soft component, even onto a black hole
of a small $\sim3\ M_{\odot}$ mass).

\begin{table}[ht]

\vspace{6mm}
\centering
{{\bf Table 2.} Results of the best-fit approximation of the
  spectrum \protect\\of \mbox{SAX\,J1818.6-1703}, averaged over
  the entire period of its activity.}\label{meansp} 

\vspace{5mm}\begin{tabular}{l|c|c|c|c|c} \hline\hline
Model\a\ &$kT,$&$\alpha$\b\ &$kT_{\rm bb},$\c\ &$L_{\rm bb},$\c\ &$\chi^2(N)$\d\ \\ 
        & keV&            & keV         &$10^{38} \ \mbox{erg s}^{-1}$&\\ \hline
PL   &              &$2.75\pm0.08$&            &           &1.02 (22)\\
PE   &$ 136\pm4 $   &$2.46\pm0.31$&            &           &1.02 (21)\\
ST   &$ 27.9\pm2.1 $&$2.51\pm0.10$&            &           &1.02 (21)\\
TB   &$ 28.8\pm1.9 $&             &            &           &1.37 (22)\\
TB+BB&$ 36.3\pm3.5 $&             &$1.7\pm0.7$ &$1.6\pm0.4$&0.75 (20)\\
ST+BB&$ 16.8\pm3.7 $&$2.08\pm0.57$&$1.7$\e\    &$1.4\pm0.5$&0.78 (20)\\ \hline
\multicolumn{6}{l}{}\\ [-3mm]
\multicolumn{6}{l}{\a\ The notation of the models is given in the text.}\\
\multicolumn{6}{l}{\b\ The photon index.}\\
\multicolumn{6}{l}{\c\ Parameters of the soft 
  radiation component: the blackbody }\\
\multicolumn{6}{l}{\ \ \ temperature and bolometric
  luminosity (for $d=8$ kpc).}\\
\multicolumn{6}{l}{\d\ The $\chi^2$ value of the best fit
  normalized to $N$ ($N$ is the number}\\
\multicolumn{6}{l}{\ \ \ of degrees of freedom).}\\
\multicolumn{6}{l}{\e\ A fixed parameter.} \\
\end{tabular}
\end{table}
\begin{table}[ht]

\vspace{6mm}
\centering
{{\bf Table 3.} Results of the best-fit approximation of the
  spectrum \protect\\ 
 of \mbox{SAX\,J1818.6-1703} at various evolutionary phases of
 the outburst}\label{evolsp}  

\vspace{5mm}\begin{tabular}{c|c|c|c|c|c} \hline\hline
Spectrum&Model&$kT,$&$\alpha$\a\ &$L_{\rm X},$\b\ &$\chi^2(N)$\c\ \\ 
      &      & keV&            &$10^{37} \ \mbox{erg s}^{-1}$& \\ \hline
A     &PL   &              &$2.74\pm0.10$&$2.37\pm0.77$ &1.03 (23)\\
      &TB   &$ 28.6\pm2.4 $&             &$2.05\pm0.18$ &0.83 (23)\\
&&&&&\\
B     &PL   &$            $&$2.55\pm0.10$&$1.98\pm0.66$ &1.25 (23)\\
      &TB   &$ 34.3\pm3.4 $&             &$1.70\pm0.15$ &0.96 (23)\\ 
&&&&&\\
C     &PL   &$            $&$2.41\pm0.11$&$2.21\pm0.79$ &1.76 (23)\\
      &TB   &$ 39.0\pm4.4 $&             &$1.88\pm0.17$ &1.04 (23)\\
&&&&&\\
D     &PL   &$            $&$3.05\pm0.18$&$0.42\pm0.21$ &0.83 (23)\\
      &TB   &$ 20.0\pm2.4 $&             &$0.36\pm0.06$ &1.23 (23)\\ \hline
\multicolumn{6}{l}{}\\ [-3mm]
\multicolumn{6}{l}{\a\ The photon index}\\
\multicolumn{6}{l}{\b\ The 20--200-keV luminosity for an assumed
  distance of $d=8$ kpc}\\ 
\multicolumn{6}{l}{\c\ The $\chi^2$ value of the best fit
  normalized to $N$ ($N$ is the number}\\ 
\multicolumn{6}{l}{\ \ \ of degrees of freedom)}\\
\end{tabular}
\end{table}

Figure~7 shows the spectral evolution of \sax\ during the
outburst under discussion. Spectra A, B, and C were measured by
the IBIS/ISGRI telescope at different evolutionary phases of the
main event (the time intervals corresponding to these phases are
indicated in the upper panel of Fig.\,5). Spectrum D was
measured during the remaining activity period of the source. The
solid lines indicate the best fit to spectrum A by the
bremsstrahlung law of an optically thin thermal plasma
($kT\simeq29$ keV). The soft component, which is absent in the
main radiation of the bursts (spectra B and C), is clearly seen
at energies below $\sim30$ keV in the spectra of the precursor
peak in the first burst (spectrum A) and the period of moderate
activity of the source (spectrum D), as well as in the average
spectrum of the source. The results of fitting the presented
spectra by a power law (PL) and the bremsstrahlung law of an
optically thin thermal plasma (TB) are given in Table~3.  We see
that in the period of moderate activity, the source had a rather
soft radiation spectrum (with a photon index of $\alpha\sim 3$);
at the onset of the main event, the spectral hardness increased
($\alpha\sim 2.7$) and continued to increase, reaching
$\alpha\sim 2.4$ in the second burst.
\begin{figure}[p]
\hspace{1.0cm}\includegraphics[width=12cm]{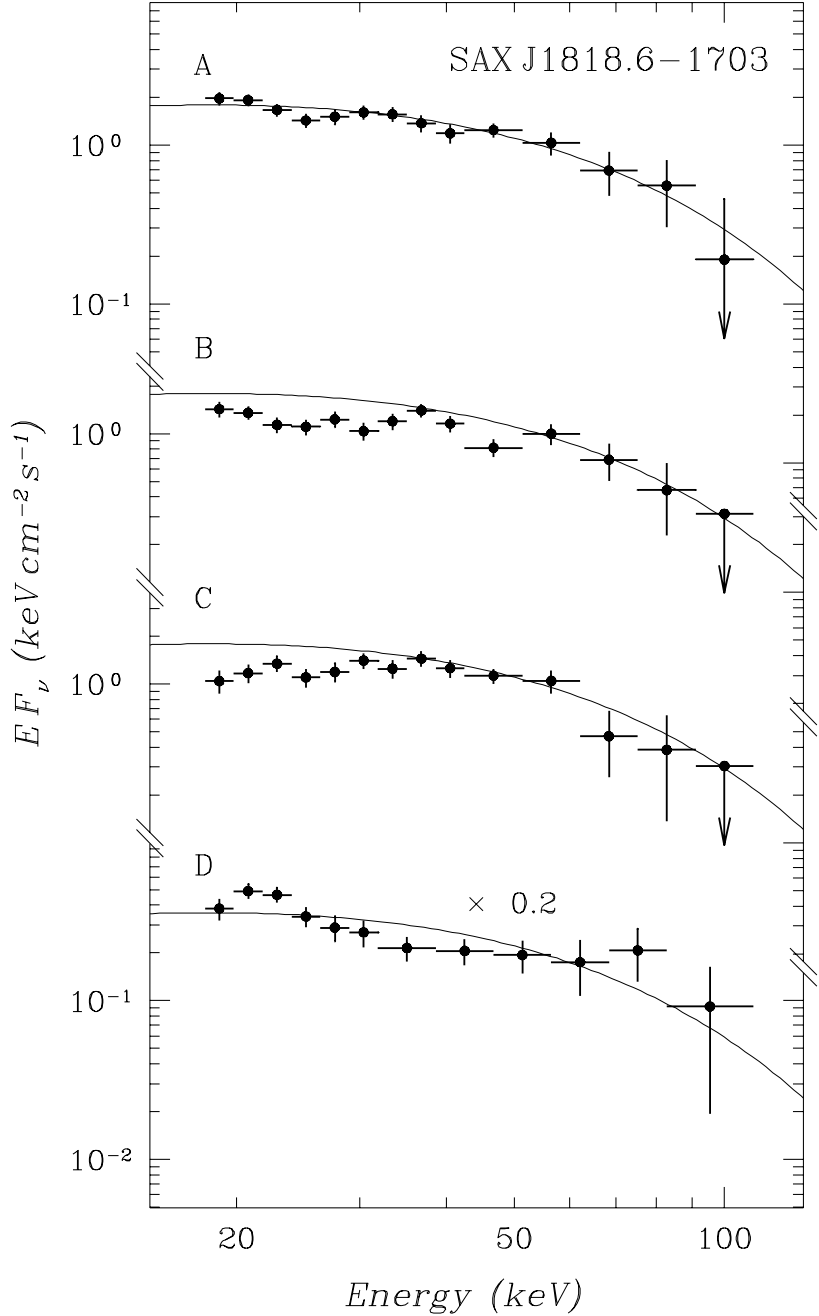}

\caption{\rm Spectral evolution of \sax\ during the hard X-ray
  outburst observed on September 9, 2003. Spectra A, B, and C
  were measured by the IBIS/ISGRI telescope at different
  evolutionary phases of the main event (see Fig.~5); spectrum D
  was measured during the remaining activity period of the
  source. The solid lines indicate the best fit to spectrum A by
  the bremsstrahlung law of an optically thin thermal plasma
  ($kT\simeq29$ keV; for spectrum D, the normalization of the
  fit was decreased by a factor of 5). The soft component, which
  is absent in the main outburst radiation (spectrum B and
  particularly spectrum C), is clearly seen at energies below
  $\sim30$ keV in the radiation spectra of the precursor peak in
  the first burst (spectrum A) and the period of moderate
  activity of the source (spectrum D).}
\end{figure}

\section*{DISCUSSION}
\noindent
Transients like \sax\ form a special, fairly representative
population among the X-ray sources discovered or recorded during
their outbursts by the \INTEGRAL\ observatory (e.g., IGR
J17544-2619, XTE J1739-302, and others). Their distinctive
feature is a very short (several hours) lifetime and a long
recurrence period (several years). Only a few such transients
(including \sax) were observed in previous experiments.

The activity of these sources could in principle be caused by
the following: (1) thermonuclear explosions on the neutron star
surface, (2) magnetic energy release in the case of a neutron
star with a very strong magnetic field, and (3) unsteady
accretion onto a neutron star or a black hole in a binary
system. The first two possibilities seem unlikely, since the
life time of such transients is much longer than the duration of
both soft gamma-ray bursts from gamma-repeaters (magnetars) and
ordinary X-ray bursts from neutron stars with a weak magnetic
field (bursters). Our results show that the bursts of \sax\ also
differ greatly from the superbursts discovered recently from
bursters (Kuulkers \etal\ 2002; in't Zand \etal\ 2004) primarily
by an increase in the hardness during the burst and by the burst
profile. Note that the energy released during the main outburst
event of \sax\ accounted for only $1/3$ of the total energy
released in the period of its activity (Table~1). This is also
difficult to explain in terms of a thermonuclear explosion on the
neutron star surface without invoking unsteady accretion
processes.

On the other hand, the lifetime of such outbursts is much shorter
than the accretion (viscous) time that characterizes the
propagation of disturbances in a standard accretion disk, 
$$t_{\rm vis}\sim\frac{2}{3\alpha}\frac{1}{\Omega_K(R)}
\left(\frac{R}{H}\right)^2\sim 1.4\ \left(\frac{R}{10^{10}\
\mbox{\rm cm}}\right)^{3/2} \left(\frac{M}{1.4\
M_{\sun}}\right)^{-1/2}\ \mbox{days}$$ (Shakura and Sunyaev
1973)\footnote{In fact, for the disk filling the Roche lobe
$t_{\rm vis}$ depends on $M$ only weakly because $R\sim M^{1/3}$
(but $t_{\rm vis}$ depends on the orbital period of the binary
and the value $R\sim10^{10}$ cm corresponds to very short
periods).}. Here, $R$ is the outer radius of the disk, $H$ is
the disk half-thickness at this radius,
$\Omega_K=(GM/R^3)^{1/2}$ is the Keplerian frequency, $M$ is the
mass of the compact object, and $\alpha\sim1$ is the viscosity
parameter. We disregarded the weak $R$-dependence of $H/R$ on
the right-hand side of this expression (in the standard disk
accretion model, $H/R\sim R^{\,1/8}$) and set $H/R\simeq 0.02$,
which corresponds to the most compact binaries with $R\sim
10^{10}$ cm (for a discussion, see Gilfanov and Arefiev
2005). We can decrease $t_{\rm vis}$ by increasing $H/R$
(advection-dominated regime), but this will cause the accretion
efficiency to decrease appreciably. At an accretion efficiency
of $\xi=1/12$ typical of black holes or compact neutron stars
($R_{\rm ns}\sim 3 R_{\rm g}$, where $R_{\rm ns}$ is the radius
of the neutron star surface, and $R_{\rm g}$ is its
gravitational radius), a compact object would accrete $M_{\rm
acc}\sim F_{\rm X}/(\xi c^2)\simeq8\times 10^{21}\ \mbox{\rm g}$
of matter for the energy release to be $F_{\rm X}\simeq
6\times10^{41}\ \mbox{erg}$ (Table~1).  The other possibility to
decrease $t_{\rm vis}$ appears in the case of unsteady accretion
from the stellar wind. In this case the radius of the forming
disk may be much smaller than the size of the binary.

The mechanisms responsible for the outbursts of such transients can
be considered in detail only after their reliable optical
identification. At present, this has not yet been done.

This work is based on the observational data obtained by the
\INTEGRAL\ observatory, an ESA satellite with instruments provided 
by ESA member states (especially France, Italy, Germany,
Switzerland, Denmark, and Spain), the Chezh Republic, and
Poland, placed in orbit by Russia and operated by the ESA with the
participation of the USA, and provided via the Russian and
European INTEGRAL Science Data Centers. We used some of the
codes developed by E.M.Churazov to analyze the data. This
study was supported by the Russian Foundation for Basic Research
(project no. 05-02-17454), the Presidium of the Russian Academy
of Sciences (the ``Nonstationary Phenomena in Astronomy'' Program),
and the Program of the Russian President for Support of Leading
Scientific Schools (project no. NSh-2083.2003.2).
 
\pagebreak   

\end{document}